\documentclass[twocolumn,showpacs,preprintnumbers,amsmath,amssymb]{revtex4-1}

\usepackage{graphicx}
\usepackage{dcolumn}
\usepackage{bm}

\begin{document}


\title{Hydrodynamic effects on concentration fluctuations in 
multicomponent membranes }

\author{Sanoop Ramachandran}
\author{Shigeyuki Komura}%
 \email{komura@tmu.ac.jp}
\affiliation{%
Department of Chemistry,
Graduate School of Science and Engineering, \\
Tokyo Metropolitan University, 
Tokyo 192-0397, Japan.
}%

\author{Kazuhiko Seki}
\affiliation{
National Institute of Advanced Industrial 
Science and Technology, \\
Ibaraki 305-8565, Japan.
}%

\author{Masayuki Imai}
\affiliation{
Department of Physics, Faculty of Science, \\ 
Ochanomizu University, Tokyo 112-0012, Japan. 
}%
\date{\today}

\begin{abstract}
We investigate the hydrodynamic effects 
on the dynamics of critical concentration fluctuations in multicomponent 
fluid membranes.
Two geometrical cases are considered; (i) confined membrane
case and (ii) supported membrane case. 
We numerically calculate the wavenumber dependence of the effective 
diffusion coefficient by changing the temperature and/or the 
thickness of the bulk fluid.
For some limiting cases, the result is compared with the 
previously obtained analytical expression.     
An analogy of the multicomponent membrane to 2D microemulsion 
is explored for the confined membrane geometry.
\end{abstract}

\maketitle

\section{Introduction}
\label{intro}

Biomembranes can be regarded as multipurpose envelopes fundamental 
to the very existence of life.
Composed of a huge variety of amphiphilic molecules, this approximately 
5~nm thick quasi-two-dimensional fluid sheet separates the inner and 
outer environments of the cells and organelles~\cite{alberts}.
Apart from delineating the cell and organelle boundaries, membranes also 
play a significant role in a variety of physiological functions such as 
trans-membrane transport of materials and cell signaling~\cite{alberts}.

Biomembranes have attracted renewed interests in the context of the 
lipid ``raft'' hypothesis proposed a little over a decade 
ago~\cite{simons-97}.
These 10--100~nm sized lipid domains with higher concentrations of 
cholesterol were proposed to play a significant role in regulating 
certain cellular functions~\cite{simons-97,edidin-03,hancock-06}.
Recent experiments using the stimulated emission depletion (STED)
microscopy on plasma membranes {\it in vivo} narrowed the size-range 
of the rafts to 10--20~nm~\cite{eggeling-09}.
Despite extensive studies in this area, the details of the 
underlying physical mechanisms leading to formation of rafts, their 
stability, and the regulation of the finite domain size remain 
elusive and controversial~\cite{munro-03,shaw-06}.

Although the biological significance of rafts is still under debate, 
there is no doubt about the rich physics and chemistry that have been 
uncovered by the membrane studies to examine the raft hypothesis.
Numerous experiments on intact cells and artificial membranes containing
saturated lipids, unsaturated lipids and cholesterol have 
demonstrated the segregation of lipids into liquid-ordered 
(${\rm L_o}$) and liquid-disordered (${\rm L_d}$) phases 
\textit{below} the miscibility transition 
temperature~\cite{veatch-03,veatch-05}.
The ${\rm L_o}$-phase is usually rich in saturated lipids and cholesterol.
Below the transition temperature, the domains undergo coarsening 
with the largest domain limited by the system size~\cite{veatch-05}.
The primary driving force for the domain coarsening is due to the positive
line tension at the domain boundaries.
Studies on the diffusion of domains~\cite{cicuta-07} and on the dynamics 
of domain coarsening~\cite{saeki-06,yanagisawa-07,veatch-07} have been 
reported.

Recently, studies on multicomponent membranes \textit{above} the 
transition temperature have also gained much attention.
As the critical point is approached from above, one observes 
composition fluctuations spanning a wide range of length and 
time scales~\cite{sengers-86}.
Veatch {\it et al.} made a notable attempt to investigate 
critical fluctuations in lipid mixtures~\cite{veatch-07}.
Deuterium NMR experiments on model ternary membranes composed of 
dioleoylphosphatidylcholine (DOPC), 
dipalmitoylphosphatidylcholine (DPPC) and cholesterol were used 
to construct the ternary phase diagram.
With the determination of the line of miscibility critical points, 
they observed that the NMR resonances were broadened in the vicinity 
of the critical points.
Such a spectral broadening was attributed to the compositional 
fluctuations in the membrane having spatial dimensions less than 
50~nm.

A more quantitative analysis of the critical fluctuations using
fluorescence microscopy was addressed by Honerkamp-Smith 
{\it et al.} for ternary mixtures of DPPC, diphytanoylphosphatidylcholine
(diPhyPC) and cholesterol~\cite{hsmith-08}.
When the critical temperature $T_{\rm c}$ is approached from above,
the correlation length diverges according to 
$\xi\approx \vert T - T_{\rm c} \vert^{-\bar{\nu}}$ where $\bar{\nu}$ 
is the critical exponent and $T$ the temperature.
(In order to prevent the confusion with the notations used later,
the critical exponents are written with a bar.)
When $T_{\rm c}$ is approached from below, on the other hand, the order 
parameter given by the difference in lipid compositions vanishes as 
$\delta \psi \approx (T_{\rm c}-T)^{\bar{\beta}}$.
From the measurements of the critical exponents $\bar{\nu}$ and 
$\bar{\beta}$, the authors concluded that the critical behavior in 
ternary membranes is in the universality class of the 2D Ising 
model~\cite{onsager-44}.
Later, it was also shown that giant plasma membrane vesicles extracted 
from that of living rat basophil leukemia cells exhibit a critical 
behavior~\cite{veatch-08}.
These experimental observations suggest that lateral heterogeneity
present in real cell membranes at physiological conditions correspond 
to critical fluctuations~\cite{hsmith-08,veatch-08}.
In other words, concentration fluctuations above the transition 
temperature (rather than below) can also be responsible for raft 
structures in cell membranes.

There have also been several theoretical works on concentration
fluctuations in multicomponent membranes.
Using renormalization group techniques, Tserkovnyak and Nelson
calculated protein diffusion in a multicomponent membrane close to 
a rigid substrate~\cite{Tserkovnyak-06}.
They pointed out that, in the vicinity of the critical point, the 
effective protein diffusion coefficient acquires a power-law behavior.
Seki {\it et al.} reported equivalent results with the use of a 
two-dimensional (2D) hydrodynamic model involving a phenomenological 
momentum decay mechanism~\cite{seki-07}.
Later Haataja showed that the effective diffusion coefficient exhibits 
a crossover from a logarithmic behavior to an algebraic dependence for 
larger length scales~\cite{haataja-09}.
In his theory, an approximate empirical relation for the diffusion 
coefficient of a moving object was employed~\cite{petrov-08}.
A rigorous hydrodynamic calculation performed by Inaura and Fujitani
arrived at similar results~\cite{inaura-08}.

\begin{figure}
\begin{center}
\includegraphics[scale=0.5]{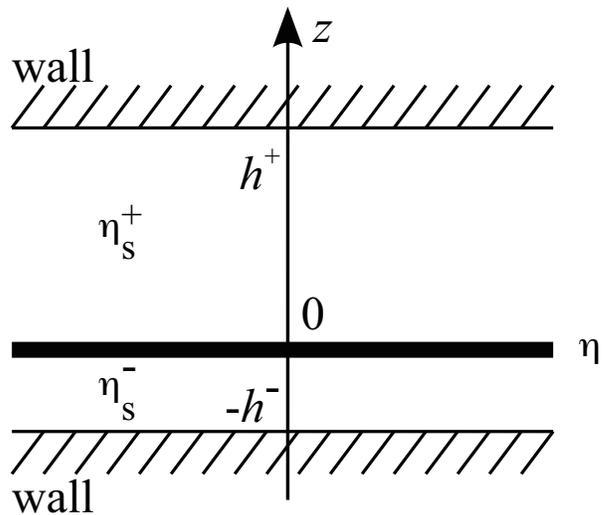}
\caption{
Schematic picture showing a planar liquid membrane having 
2D viscosity $\eta$ located at $z=0$.
It is sandwiched by a solvent of 3D viscosity $\eta_{\rm s}^{\pm}$.
Two impenetrable walls are located at $z=\pm h^\pm$ bounding the 
solvent.
}
\label{fig:membrane-wall}
\end{center}
\end{figure}

The present article uses the idea of critical phenomena to 
calculate the effective diffusion coefficient in multicomponent 
lipid membranes.
Based on the time-dependent Ginzburg-Landau approach with full 
hydrodynamics, we calculate in particular the decay rate of the 
concentration fluctuations occurring in membranes.    
We deal with the case where the membrane is surrounded by a bulk
solvent and two walls as depicted in Fig.~\ref{fig:membrane-wall}.
Such a situation is worth considering because biological membranes 
interact strongly with other cells, substrates or even the underlying
cytoskeleton which affects the structural and transport properties 
of the membrane~\cite{simons-04}.
We note that effects of the substrates on the diffusion coefficient 
of proteins have been investigated before~\cite{evans-88,stone-98}.
Two surrounding geometries of the membrane are discussed; (i) confined 
membrane and (ii) supported membrane.  
We also study the situation when the multicomponent membranes form 
2D microemulsions~\cite{gompper-schick}.
This interesting viewpoint is motivated by a recent 
work which predicts the reduction of the line tension in membranes 
containing saturated, unsaturated and hybrid lipids (one tail 
saturated and the other unsaturated)~\cite{brewster-09,brewster-10}.
Based on chain entropy arguments, they proposed that hybrid lipids 
in 2D play an equivalent role to surfactant molecules in 
three-dimensions (3D).
Hence we shall explore the concentration fluctuations in 2D 
microemulsion.

This paper is organized as follows.
In Section~\ref{membrane}, we first obtain the membrane mobility 
tensors that are used in the subsequent calculations.
In Section~\ref{concfluct}, the decay rates of concentration
fluctuations are analyzed for confined and supported membrane cases.
Section~\ref{microemul} deals with the microemulsion picture of the 
multicomponent membranes. 
Finally we close with some discussions in Section~\ref{discussion}.

\section{Membrane hydrodynamics}
\label{membrane}

In this section, we start with the membrane equations of motion 
as well as the requisite membrane mobility tensors for the two 
geometries.
The membrane is assumed to be an infinite planar sheet of liquid
and its out-of-plane fluctuations are totally neglected, which is justified 
for typical bending rigidities of bilayers.
The liquid membrane, fixed in the $xy$-plane at $z=0$,
is embedded in a bulk fluid such as water or solvent that is further 
bounded by hard walls as shown in Fig.~\ref{fig:membrane-wall}.
The upper ($z>0$) and the lower ($z<0$) fluid regions are denoted
by $``+"$ and $``-"$, respectively. 
Since the 3D viscosities of the upper and the lower 
solvent can be different, we denote them as $\eta_{\rm s}^{\pm}$. 
Consider the situation in which impenetrable walls are located 
at $z=\pm h^{\pm}$, where $h^+$ and $h^-$ can also be different in 
general.

Let ${\bf v}({\bf r})$ be the 2D velocity of the membrane fluid. 
The 2D vector ${\bf r}=(x,y)$ represents a point in the plane 
of the membrane which 
is assumed to be incompressible  
\begin{equation}
\nabla \cdot {\bf v} = 0. 
\label{eqn:2Dcompress}
\end{equation}
Here $\nabla$ is a 2D differential operator.
We work in the low-Reynolds number regime of the membrane hydrodynamics
so that the inertial effects can be neglected.
This allows us to use the 2D Stokes equation given by 
\begin{equation}
\eta \nabla^2 {\bf v} - \nabla p + {\bf f}_{\rm s} + \bf{F} =0,
\label{eqn:2Dstokes}
\end{equation}
where $\eta$ is the 2D membrane viscosity,
$p$ the 2D in-plane pressure,
${\bf f}_{\rm s}$ the force exerted on the membrane by the 
surrounding fluid (``s'' stands for the solvent), 
and ${\bf F}$ is any other force acting on the membrane.

Once we know ${\bf f}_{\rm s}$, the membrane velocity can be obtained 
from eqn~(\ref{eqn:2Dstokes}) as  
\begin{equation}
{\bf v}[{\bf q}]
= {{\bf G}}[{\bf q}] \cdot {\bf F}[{\bf q}],
\label{eqn:membvelFT}
\end{equation}
where ${\bf v}[{\bf q}]$, ${\bf G}[{\bf q}]$ and ${\bf F}[{\bf q}]$ 
(${\bf q}=(q_x,q_y)$) are the Fourier components of 
${\bf v}({\bf r})$, ${\bf G}({\bf r})$ and ${\bf F}({\bf r})$ defined by
\begin{equation}
{\bf v}({\bf r}) = 
\int \frac{{\rm d} {\bf q}}{(2\pi)^2}
{\bf v}[{\bf q}]
\exp ( i {\bf q}\cdot {\bf r}),
\label{eqn:FT}
\end{equation}
\begin{equation}
{\bf G}({\bf r}) = 
\int \frac{{\rm d} {\bf q}}{(2\pi)^2}
{\bf G}[{\bf q}]
\exp ( i {\bf q}\cdot {\bf r}),
\label{eqn:mob}
\end{equation}
and
\begin{equation}
{\bf F}({\bf r}) = 
\int \frac{{\rm d} {\bf q}}{(2\pi)^2}
{\bf F}[{\bf q}]
\exp ( i {\bf q}\cdot {\bf r}),
\end{equation} 
respectively. 
Using the stick boundary conditions at $z=0$ and $z=\pm h^\pm$, we 
can solve the hydrodynamic equations to obtain 
${\bf f}_{\rm s}$~\cite{sanoop-poly-10}.
Its perpendicular component with respect to ${\bf q}$ is given by
\begin{align}
f_{\rm s\perp}[{\bf q}] 
= - \eta_{\rm s}^+ v_\perp q \coth(qh^+)
- \eta_{\rm s}^- v_\perp q \coth(qh^-),
\end{align}
where $q=\vert {\bf q} \vert$ and $v_\perp$ is also the 
perpendicular component of ${\bf v}$ with respect to ${\bf q}$.
On the other hand, one can show that the parallel component of 
${\bf f}_{\rm s}$ is zero.
After some calculations, the components of the mobility tensor 
${\bf G}[{\bf q}]$ finally becomes
\begin{align}
G_{\alpha\beta}[{\bf q}] = &  
\frac{1}{\eta q^2 + q[ \eta_{\rm s}^+ \coth(qh^+)
+\eta_{\rm s}^-\coth(qh^-)]} 
\nonumber\\
& \times \left( \delta_{\alpha\beta} - \frac{q_\alpha q_\beta}{q^2} \right),
\label{eqn:genoseen}
\end{align}
with $\alpha, \beta = x,y$.
The full details of the calculation are given in the separate
article by Ramachandran {\it et al.}~\cite{sanoop-poly-10}.

As in ref.~\cite{inaura-08}, we first consider 
the case when the two walls are located at equal distances from 
the membrane, i.e., $h^+=h^-=h$.
Then the above mobility tensor simplifies to
\begin{equation}
G_{\alpha\beta}[{\bf q}] =  \frac{1}{\eta [ q^2 + \nu q \coth(qh)]} 
\left( \delta_{\alpha\beta} - \frac{q_\alpha q_\beta}{q^2} \right),
\label{eqn:cothoseen}
\end{equation}
where $\nu \equiv 2 \eta_{\rm s} / \eta$ with 
$\eta_{\rm s} = (\eta_{\rm s}^{+}+\eta_{\rm s}^{-})/2$.
This expression, taken in the limit of large $h$, has been 
used previously in calculating the correlated diffusion coefficients
of particles embedded in a membrane~\cite{oppenheimer-09}.

For supported membranes, $h^+$ is infinitely large while $h^-$ is
finite when compared to the membrane thickness.
In this case, eqn~(\ref{eqn:genoseen}) reduces to~\cite{sanoop-poly-10}
\begin{equation}
G_{\alpha\beta}[{\bf q}] =  \frac{1}{\eta [ q^2 + \nu q(1+ \coth(qh^-))/2]} 
\left( \delta_{\alpha\beta} - \frac{q_\alpha q_\beta}{q^2} \right).
\label{eqn:sup-oseen}
\end{equation}
This equation has been used for the investigation of the 
correlated dynamics of inclusions in a supported 
membrane~\cite{oppenheimer-10}.

\section{Dynamics of concentration fluctuations}
\label{concfluct}

In order to discuss the dynamics of concentration fluctuations above
the transition temperature, we closely follow the formalism used 
in ref.~\cite{nonomura-99}.
Here we extend our previous work in ref.~\cite{seki-07} for membranes
embedded in a solvent confined by two walls (confined membrane) or 
supported on a substrate (supported membrane).

Consider a two-component fluid membrane composed of lipid A 
and lipid B whose local area fractions are denoted by 
$\phi_{\rm A} ({\bf r})$ and $\phi_{\rm B} ({\bf r})$, respectively.
Since the relation $\phi_{\rm A} ({\bf r}) + \phi_{\rm B} ({\bf r})=1$
holds, we introduce a new variable defined by
$\psi ({\bf r}) \equiv \phi_{\rm A} ({\bf r}) - \phi_{\rm B} ({\bf r})$.
The simplest form of the free energy functional ${\mathcal F}\{\psi\}$
describing the fluctuation around the homogeneous state is
\begin{equation}
{\mathcal F}\{\psi\} = \int {\rm d}{\bf r} 
\left[ \frac{a}{2}\psi^2 + \frac{c}{2} (\nabla \psi)^2 - \mu \psi \right],
\label{eqn:freeE}
\end{equation}
where $a >0$ is proportional to the temperature difference from the
critical temperature, $c>0$ is related to the line tension and
$\mu$ is the chemical potential.

The time evolution of concentration in the presence of hydrodynamic
flow is given by the time-dependent Ginzburg-Landau equation for
a conserved order parameter~\cite{nonomura-99}
\begin{equation}
\frac{\partial \psi}{\partial t} + \nabla \cdot ({\bf v}\psi) 
= L \nabla^2\frac{\delta {\mathcal F}}{\delta \psi},
\label{sec:concfluct:eqn4}
\end{equation}
where $L$ is the kinetic coefficient.
In the membrane hydrodynamic equation (\ref{eqn:2Dstokes}), on 
the other hand, we need to incorporate the thermodynamic force due 
to the concentration fluctuations. 
Hence we have 
\begin{equation}
\bf{F} = -\psi \nabla \frac{\delta {\mathcal F}}{\delta \psi}.
\end{equation}

We implicitly assumed that the relaxation of the velocity ${\bf v}$ 
is much faster than that of concentration $\psi$~\cite{seki-07}.
The membrane velocity can be formally solved using the appropriate
2D mobility tensor $G_{\alpha\beta}({\bf r},{\bf r}')$ 
derived in the previous section,
\begin{equation}
v_\alpha({\bf r},t) = 
\int {\rm d}{\bf r}' 
G_{\alpha \beta}({\bf r},{\bf r}')
(\nabla_\beta'\psi) \frac{\delta {\mathcal F}}{\delta \psi ({\bf r}')}.
\label{sec:concfluct:eqn5}
\end{equation}
Since our interest is in the concentration fluctuations around
the homogeneous state, we define 
$\delta \psi({\bf r},t) =  \psi({\bf r},t) - \bar{\psi}$,
where the bar indicates the spatial average.
The free energy functional expanded in powers of $\delta \psi$
becomes,
\begin{equation}
{\mathcal F}\{\delta\psi\} = \int {\rm d}{\bf r} 
\left[ \frac{a}{2}(\delta \psi)^2 + \frac{c}{2} (\nabla \delta \psi)^2 \right].
\end{equation}
Substituting eqn~(\ref{sec:concfluct:eqn5}) into 
eqn~(\ref{sec:concfluct:eqn4}), we get
\begin{align}
&\frac{\partial \delta \psi ({\bf r},t)}{\partial t} = 
L\nabla^2 \frac{\delta {\mathcal F}}{\delta (\delta \psi)}  
\nonumber\\
&-\int {\rm d}{\bf r}' 
(\nabla_\alpha \delta\psi({\bf r}))
G_{\alpha \beta}({\bf r},{\bf r}')
(\nabla_\beta' \delta\psi({\bf r}')) 
\frac{\delta {\mathcal F}}{\delta (\delta \psi ({\bf r}'))}.
\end{align}

We now consider the dynamics of the time-correlation function defined by
\begin{equation}
S({\bf r},t) = \langle \delta \psi({\bf r}_1,t) \delta \psi({\bf r}_2,0) 
\rangle, 
\end{equation}
where ${\bf r} = {\bf r}_2 - {\bf r}_1$. 
Within the factorization approximation~\cite{nonomura-99},
the spatial Fourier transform of $S({\bf r},t)$
defined by
\begin{equation}
S({\bf r},t) = 
\int \frac{{\rm d} {\bf q}}{(2\pi)^2}
S[{\bf q},t]
\exp ( i {\bf q}\cdot {\bf r}),
\end{equation}
satisfies the following equation
\begin{equation}
\frac{\partial S[{\bf q},t]}{\partial t} =
-\left( \Gamma^{(1)}[{\bf q}] 
+ \Gamma^{(2)}[{\bf q}] \right) S[{\bf q},t].
\label{eqn:tcf}
\end{equation} 
In the above, the first term $\Gamma^{(1)}[{\bf q}]$ denotes the 
van Hove part of the relaxation rate given by
\begin{equation}
\Gamma^{(1)}[{\bf q}] =  Lk_{\rm B}Tq^2 \chi^{-1}[{\bf q}].
\end{equation}
Here the static correlation function is defined by
\begin{equation}
\chi[{\bf q}]
= \langle \delta \psi[{\bf q}] \delta \psi[-{\bf q}] \rangle
= \frac{k_{\rm B} T}{c(q^2 + \xi^{-2})},
\end{equation}
where $\xi \equiv (c/a)^{1/2}$ is the correlation length,
$k_{\rm B}$ the Boltzmann constant,
and $T$ the temperature.

As for the second term in eqn~(\ref{eqn:tcf}), $\Gamma^{(2)}[{\bf q}]$ 
denotes the hydrodynamic part of the decay rate 
\begin{align}
\Gamma^{(2)}&[{\bf q}]=  \frac{1}{\chi[{\bf q}]}
\int \frac{{\rm d} {\bf p}}{(2\pi)^2}
q_\alpha G_{\alpha\beta}[{\bf p}] q_\beta \chi[{\bf q}+{\bf p}],
\label{eqn:gammahydro}
\end{align}
where either eqn~(\ref{eqn:cothoseen}) or eqn~(\ref{eqn:sup-oseen}) 
will be used for the mobility tensor $G_{\alpha\beta}$.

\subsection{Confined membrane}

\begin{figure}
\begin{center}
\includegraphics[scale=0.47]{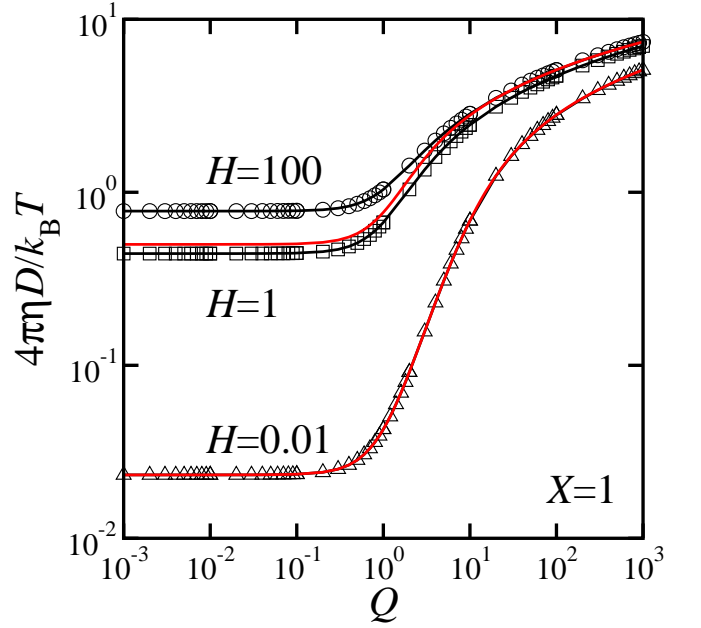}
\caption{
Scaled effective diffusion coefficient $D$ as a function of $Q$
for $H=0.01, 1, 100$ when $X=1$ for the confined membrane case.
The numerically calculated points are joined by lines for clarity.
The solid red lines are from the analytical expression given
in eqn~(\ref{eqn:sk}) obtained in the limit of small $H$.
}
\label{fig:DvsQ-Coth-N1}
\end{center}
\end{figure}

First we consider the situation in which the membrane is confined by 
two walls which are located at equidistant $h$ from the membrane.
The appropriate form of the mobility tensor is given by 
eqn~(\ref{eqn:cothoseen}).
Substituting it into eqn~(\ref{eqn:gammahydro}), the hydrodynamic 
part of the decay rate is expressed as 
\begin{align}
& \Gamma^{(2)}[{\bf q}]=
 \frac{k_{\rm B}T}{\eta \chi[{\bf q}]}
\int \frac{{\rm d} {\bf p}}{(2\pi)^2}\nonumber\\
&\times \frac{\chi[{\bf p}]}
{ |{\bf p} - {\bf q}|^2 + \nu |{\bf p} - {\bf q}| \coth(|{\bf p} - {\bf q}|h)}
\frac{ q^2 p^2  - ({\bf q} \cdot {\bf p})^2}{|{\bf p} - {\bf q}|^2}.
\end{align}
We introduce an effective diffusion coefficient $D[{\bf q}]$ 
(due only to the hydrodynamic effect) defined by
\begin{equation}
\Gamma^{(2)}[{\bf q}]= q^2 D[{\bf q}].
\end{equation}
In order to deal with dimensionless quantities, we rescale all the 
lengths by the hydrodynamic screening length 
$\nu^{-1}=\eta/(2\eta_{\rm s})$ such that 
$P \equiv p /\nu$, $Q\equiv q /\nu$,  
$X\equiv \xi\nu$ and  $H \equiv h\nu$.
Then $D[{\bf q}]$ can be rewritten as
\begin{align}
& D[Q;X,H]= \frac{k_{\rm B}T}{4\pi^2\eta}(1+Q^2X^2)
\nonumber\\
& \times \int_0^\infty {\rm d}P \int_{0}^{2\pi} {\rm d}\theta
\frac{ P^3 \sin^2\theta}
{(1+P^2X^2)[ G^2+ G^{3/2}\coth(\sqrt{G}H)]},
\label{eqn:cothDintegral}
\end{align}
with $G=P^2+Q^2-2PQ\cos\theta$.
Since this integral cannot be performed analytically, we 
evaluate it numerically.
We explore the dependencies of $D$ on the variable $Q$, and the 
parameters $X$ and $H$.

\begin{figure}
\begin{center}
\includegraphics[scale=0.47]{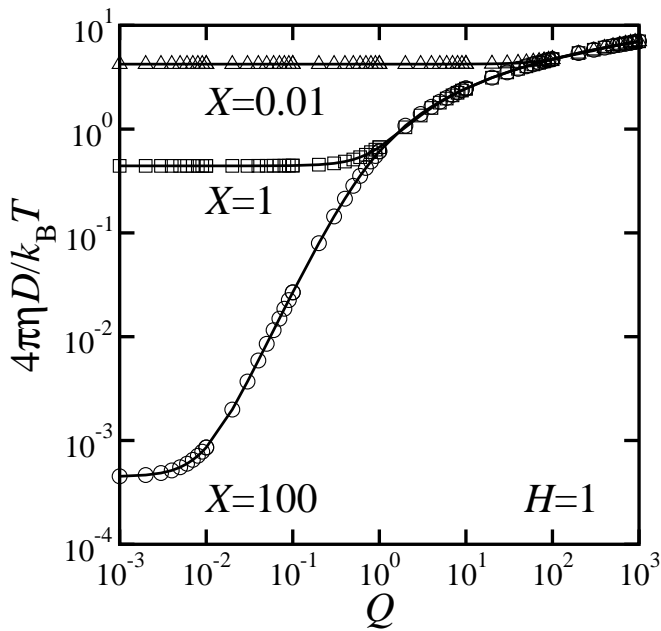}
\caption{
Scaled effective diffusion coefficient $D$ as a function of $Q$
for $X=0.01, 1, 100$ when $H=1$ for the confined membrane case.
}
\label{fig:DvsQ-Coth-H1}
\end{center}
\end{figure}

In Fig.~\ref{fig:DvsQ-Coth-N1}, we plot the diffusion coefficient 
$D$ (scaled by $k_{\rm B}T/4\pi\eta$) as a function of dimensionless
wavenumber $Q$ for different solvent thickness $H$ while the 
correlation length is fixed to $X=1$ (i.e., fixed temperature). 
In the limit of $Q \ll 1$, $D$ is almost a constant.
The calculated $D$ starts to increase around $Q\approx 1$ and a 
logarithmic behavior (extracted via numerical fitting) 
is seen for $Q\gg1$.
This trend has been previously reported by
Seki {\it et al.}~\cite{seki-07} or Inaura and Fujitani~\cite{inaura-08}.
When $H$ is small such as $H=0.01$, the value of $D$ 
decreases by one order of magnitude compared to $H=100$.
Figure~\ref{fig:DvsQ-Coth-H1} shows the diffusion coefficient $D$ 
as a function of wave number $Q$ for different $X$ (i.e., different 
temperature) while the solvent height is fixed to $H=1$.
Again, $D$ is nearly constant for $Q \ll 1$, and follows an $S$-shaped 
curve with increasing $Q$.
Finally, a logarithmic dependence is observed from numerical fitting. 
We note that the above logarithmic behavior for $Q \gg 1$ is in 
contrast to that of 3D critical fluids as given by the Kawasaki 
function which increases linearly with $q$~\cite{kawasaki-70}.

\begin{figure}
\begin{center}
\includegraphics[scale=0.47]{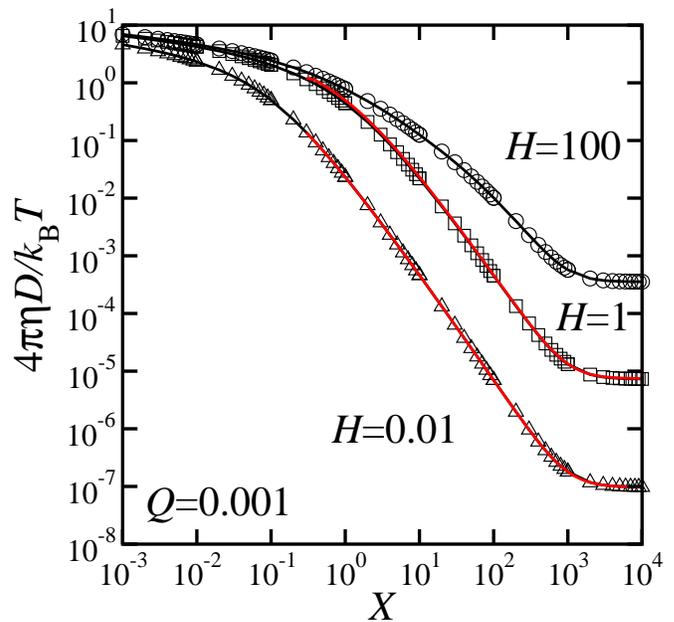}
\caption{
Scaled effective diffusion coefficient $D$ as a function of $X$
for $H=0.01, 1, 100$ when $Q=10^{-3}$ for the confined membrane case.
The solid red lines are from the analytical expression given
in eqn~(\ref{eqn:sk}) obtained in the limit of small $H$.
}
\label{fig:DvsN-Coth}
\end{center}
\end{figure}

\begin{figure}
\begin{center}
\includegraphics[scale=0.47]{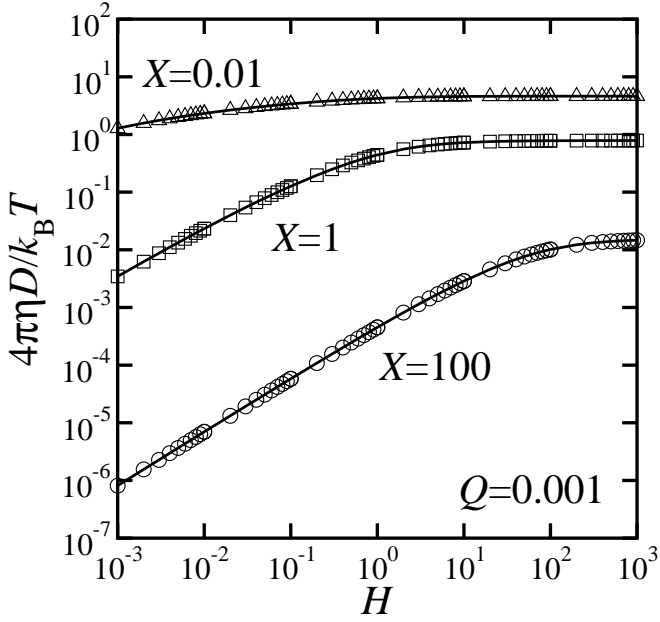}
\caption{
Scaled effective diffusion coefficient $D$ as a function of $H$
for $X=0.01, 1, 100$ when $Q=10^{-3}$ for the confined membrane case.
}
\label{fig:DvsH-Coth}
\end{center}
\end{figure}

In Fig.~\ref{fig:DvsN-Coth}, we explore the effect of the correlation 
length $X$ on $D$ for different values of $H$ when $Q=10^{-3}$.
The quantity $X$ measures an effective size of the correlated
region formed transiently in the membrane due to thermal fluctuations.
When $X \ll 1$, the diffusion coefficient $D$ decreases only 
logarithmically, which is typical for a pure 2D 
system~\cite{saffman-75,saffman-76,hughes-81}.
When $X\gg1$, on the other hand, the behavior of $D$ depends on the 
value of $H$.
The proximity to the walls results in a loss of 
momentum from the membrane~\cite{seki-07,sanoop-poly-10,diamant-09b}.
This leads to a rapid suppression of the velocity field within the 
membrane such that the concentration fluctuations decay slowly.
Consequently, the values of $D$ are lower for smaller $H$. 
The flattening of the curves for large $X$ is due to the dominance 
of the $X^2$ terms in the numerator and denominator of 
eqn~(\ref{eqn:cothDintegral}).
In Fig.~\ref{fig:DvsH-Coth}, we plot $D$ as a function of $H$ for 
different values of $X$ when $Q=10^{-3}$.
In general, there is a monotonic increase in $D$ with larger $H$ 
followed by a saturation to a constant value.
We see that the effect of $H$ is most prominent for large $X$
(close to the critical point), while there is only a weak dependence 
for $X \ll 1$ (far from the critical point).
The former reflects the fact that the membrane fluid is affected 
by the the outer environment when the correlation length 
$\xi$ is larger than the hydrodynamic screening length 
$\nu^{-1}$ i.e, $X\gg1$~\cite{sanoop-poly-10}.
For correlation lengths smaller than the $\nu^{-1}$, the outer 
environment surrounding the membrane is less important.
In this situation, the system behaves essentially as a pure 2D one.
When the correlation length becomes larger than $\nu^{-1}$, as is 
the case near the critical point, the outer environment significantly 
affects membrane dynamics.

For small $h$, the $\nu q \coth(qh)$ term in the mobility tensor
eqn~(\ref{eqn:cothoseen}) can be replaced by a constant
$\nu/h$. 
In this case, Seki {\it et al.} obtained an analytical expression for 
the effective diffusion coefficient~\cite{seki-07}.
Their result in terms of the dimensionless quantities $Q$, $X$ and
$H$ can be reproduced as
\begin{align}
&D[Q;X,H] = \frac{k_{\rm B}T}{4\pi\eta} \frac{1+Q^2X^2}{2Q^2X^2}
\left[ -\ln \left( \frac{X}{\sqrt{H}} \right) \right.
\nonumber\\
& 
+ \frac{H}{X^2}(1+Q^2X^2)
\ln\left( \frac{X}{\sqrt{H}(1+Q^2X^2)} \right)
\nonumber\\
&
\left.
+\frac{H\Omega}{2X^2}
\ln\left( \frac{ Q_+^4 + Q_-^2 + Q_+^2 \Omega}{\Omega-Q_-^2-1} \right)
\right],
\label{eqn:sk}
\end{align}
where
\begin{equation}
\Omega = \sqrt{\left( Q^2 X^2 + X^2/H -1 \right)^2 
+4 Q^2 X^2},
\end{equation}
and
\begin{equation}
Q_\pm = \sqrt{Q^2 X^2 \pm X^2/H}.
\end{equation}
Equation (\ref{eqn:sk}) is plotted using red curves in 
Fig.~\ref{fig:DvsQ-Coth-N1} for $H=0.01$ and $1$ with
$X=1$.
For $H=0.01$, the analytical and numerical data coincide
giving credence to accuracy of the numerical solutions.
It is seen that even for $H=1$ the agreement is still acceptable.
For $H=100$, however, a significant deviation is observed (not shown), 
which is expected as this limit is beyond the valid range of 
eqn~(\ref{eqn:sk}).

The red curves in Fig.~\ref{fig:DvsN-Coth} also represent the analytical
result of eqn~(\ref{eqn:sk}).
It is seen that the analytical and the numerical data points almost 
coincide for $H=0.01$ and $H=1$.
For $H=100$, there is significant deviation from the numerical data 
(not shown).
However, the agreement between the numerical result and the analytical 
expression is beyond the expected range of $H \ll 1$ and reaches up to 
$H \approx 1$, as pointed out by Stone and Ajdari~\cite{stone-98}.
Hence eqn~(\ref{eqn:sk}) can be useful in analyzing the experimental
data in many situations.

\subsection{Supported membrane}

\begin{figure}
\begin{center}
\includegraphics[scale=0.47]{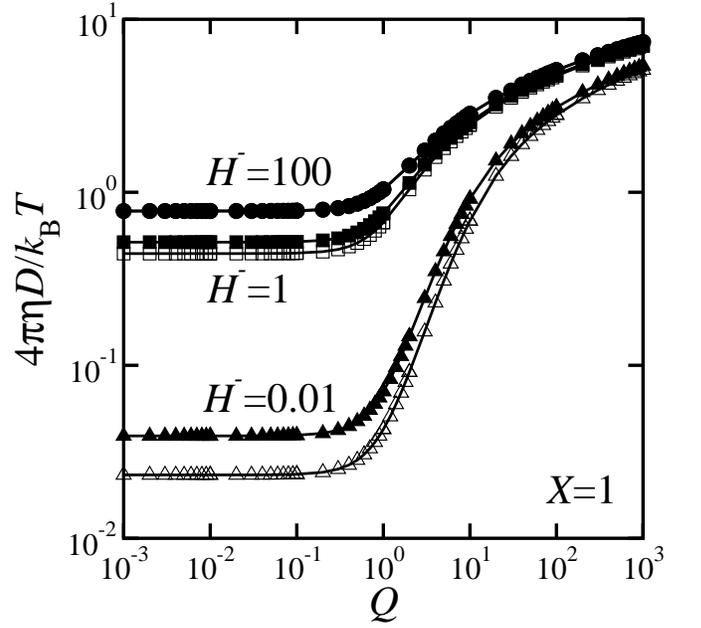}
\caption{
Scaled effective diffusion coefficient $D$ as a function of $Q$
for $H^-=0.01, 1, 100$ when $X=1$ for the supported membrane case
plotted with filled symbols.
For comparison, the data in Fig.~\ref{fig:DvsQ-Coth-N1} are 
also plotted with open symbols.
}
\label{fig:supp-DvsQ-Coth-N1}
\end{center}
\end{figure}

Apart from studying membrane dynamics on vesicles, experiments are 
also conducted on supported membranes which have a benefit of avoiding
curvature effects~\cite{kaizuka-04}.
We proceed now to calculate concentration fluctuations for the supported
membrane case.
As given by eqn~(\ref{eqn:sup-oseen}), the mobility tensor of a 
supported membrane is slightly modified from that of a confined 
membrane. 
By substituting eqn~(\ref{eqn:sup-oseen}) into 
eqn~(\ref{eqn:gammahydro}), we obtain the expression for the effective 
diffusion coefficient as
\begin{align}
&D[Q;X,H^-]= \frac{k_{\rm B}T}{4\pi^2\eta} (1+Q^2X^2)
\int_0^\infty {\rm d}P \int_{0}^{2\pi} {\rm d}\theta
\nonumber\\
& \times \frac{ P^3 \sin^2\theta}
{(1+P^2X^2)[ G^2+ G^{3/2}(1+\coth(\sqrt{G}H^-))/2]},
\end{align}
where $H^-\equiv h^-\nu$ and $G=P^2+Q^2-2PQ\cos\theta$ as before.

Performing numerical integrations, we plot in 
Fig.~\ref{fig:supp-DvsQ-Coth-N1} the effective diffusion 
coefficient $D$ as a function of $Q$ for different values
of $H^-$ when $X=1$ (closed symbols).
In general, the behavior of $D$ is similar to that of the confined 
membrane case.
When $H^- \ll 1$, the calculated $D$ is slightly larger than the 
confined membrane compared with the same value of $H$.
This is because the supported membrane is subjected to only one wall,
while the confined membrane is sandwiched by two walls on both sides.
The confined and the supported membranes show an almost identical 
behavior when both $H$ and $H^-$ are large enough, as it should be.
The other dependencies of $D$ are similar to the confined membrane 
case except for a slight increase at small $H^-$ values.

\section{Membrane as a 2D microemulsion}
\label{microemul}

The role of surfactant molecules in 3D microemulsions is to reduce 
the surface tension at the interface between oil and water.
In an analogy to 3D microemulsions, hybrid lipids (one chain 
unsaturated and the other saturated) act as  lineactant molecules 
which stabilize finite sized domains in 2D. 
In other words, hybrid lipids play a similar role to surfactant 
molecules at the interface between ${\rm L_o}$ and ${\rm L_d}$ 
domains.
It should be also noticed that hybrid lipids form a major percentage 
of all naturally existing lipids~\cite{vandeenen-71,vanmeer-08}.
Based on a simple model of hybrid lipids, Brewster {\it et al.} 
showed that finite sized domains can be formed in 
equilibrium~\cite{brewster-09,brewster-10}.
A subsequent model predicted stabilized domains even in a system of 
saturated/hybrid/cholesterol lipid membranes~\cite{yamamoto-10}.
Being motivated by this idea, we calculate the decay rate of concentration 
fluctuations when the free energy of the multicomponent membrane 
has the form of a 2D microemulsion. 
Here we consider only the confined membrane geometry, and use 
eqn~(\ref{eqn:cothoseen}) for the mobility tensor.

The free energy functional for a microemulsion includes a higher 
order derivative term and is expressed in terms of $\delta \psi$ 
as~\cite{gompper-schick}
\begin{equation}
{\mathcal F}_{\rm ME}\{\delta\psi\} = \int {\rm d}{\bf r} 
\left[ 
\frac{a}{2}(\delta \psi)^2 + 
\frac{c}{2} (\nabla \delta \psi)^2 +
\frac{g}{2} (\nabla^2 \delta \psi)^2 
\right],
\end{equation}
with $a,g>0$ and $c<0$.
The negative value of $c$ creates 2D interfaces, while the term 
with positive $g$ is a stabilizing term.
This form of the free energy has been used previously to study
coupled modulated bilayers~\cite{hirose-09}.
As in the previous section, the decay rate of the correlation 
function can be split into two parts.
First, the van Hove part becomes now
\begin{equation}
\Gamma^{(1)}_{\rm ME}[{\bf q}] = L k_{\rm B} T 
q^2 \chi^{-1}_{\rm ME}[{\bf q}],
\label{eq:mevanhove}
\end{equation}
where $L$ is the kinetic coefficient assumed to be same 
as before, and the static correlation function 
$\chi_{\rm ME}[{\bf q}]$ is~\cite{strey-87}  
\begin{equation}
\chi_{\rm ME}[{\bf q}]
=\frac{k_{\rm B} T}{gq^4+cq^2+a}.
\end{equation}
By defining
\begin{equation}
q_0^2 =  -\frac{c}{2g}, 
\end{equation}
\begin{equation}
\sigma^4 = \frac{a}{g} - \left(\frac{c}{2g}\right)^2,
\end{equation}
the static correlation function can be also written as
\begin{equation}
\chi_{\rm ME}[{\bf q}]
=\frac{k_{\rm B} T}{g\left[ (q^2 - q_0^2)^2 + \sigma^4 \right]}.
\label{eqn:statcorfn}
\end{equation}
On plotting $\chi_{\rm ME}$ as a function of $q$, a peak appears 
at $q=q_0$ followed by a $q^{-4}$-decay. 
The width of the peak is given by $\sigma$, and
a lamellar phase appears when $\sigma=0$.
Notice that $c=0$ is called the Lifshitz point at which the peak 
occurs for $q=0$~\cite{gompper-90}.
Using the form of eqn~(\ref{eqn:statcorfn}), we 
can rewrite eqn~(\ref{eq:mevanhove}) as
\begin{equation}
\Gamma^{(1)}_{\rm ME}[{\bf q}] =  L g q^2 
\left[ (q^2 - q_0^2)^2 + \sigma^4 \right].
\end{equation}

\begin{figure}
\begin{center}
\includegraphics[scale=0.47]{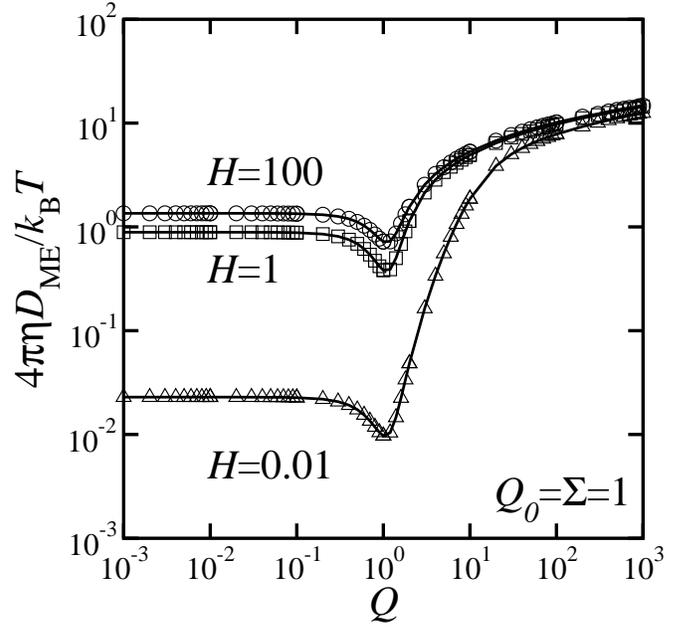}
\caption{
Scaled effective diffusion coefficient $D_{\rm ME}$ as a function 
of $Q$ for $H=0.01, 1, 100$ when $Q_0=\Sigma=1$ for the confined 
membrane case.
}
\label{fig:mDvsQ-b1N1}
\end{center}
\end{figure}

As in the previous section, we next write the hydrodynamic part of the 
decay rate in terms of the effective diffusion coefficient 
$D_{\rm ME}[{\bf q}]$ as 
\begin{equation}
\Gamma^{(2)}_{\rm ME}[{\bf q}]=q^2 D_{\rm ME}[{\bf q}].
\end{equation}
Using eqn~(\ref{eqn:cothoseen}) for the mobility tensor, 
we obtain $D_{\rm ME}$ as 
\begin{align}
&D_{\rm ME}[Q;Q_0,\Sigma,H] = 
\frac{k_{\rm B}T}{4\pi^2 \eta}  
[ (Q^2 - Q_0^2)^2 + \Sigma^4 ]
\int_0^\infty {\rm d}P
\nonumber\\
& \times 
\int_{0}^{2\pi} {\rm d}\theta
\frac{P^3 \sin^2\theta}
{[(P^2 - Q_0^2)^2 + \Sigma^4 ] 
[G^2 + G^{3/2}\coth(\sqrt{G}H)]},
\end{align}
where $P \equiv p/\nu$, $Q\equiv q/\nu$, 
$Q_0 \equiv q_0/\nu$, $\Sigma \equiv \sigma/ \nu$, 
$H\equiv h\nu$, and $G = P^2 + Q^2 - 2PQ \cos\theta$.

In Fig.~\ref{fig:mDvsQ-b1N1}, we plot $D_{\rm ME}$ as a function 
of $Q$ for different values of $H$ when $Q_0=\Sigma=1$ are fixed.
When $Q \ll 1$, $D$ shows a constant value.
We also observe the 2D characteristic of logarithmic behavior
of $D$ for $Q\gg1$.
For $Q\ll1$, the effect of the outer environment is felt with the 
suppression of the diffusion coefficient with smaller $H$.
The curves almost overlap when $Q\gg1$ indicating the negligible 
effect of the outer environment at large wave numbers.
An interesting feature of $D_{\rm ME}$ is the dip occurring at 
$Q\approx Q_0$ which does not exist for binary critical fluids. 
This can be attributed to the peak at $q=q_0$ in 
$\chi_{\rm ME}[{\bf q}]$~\cite{nonomura-99,strey-87,hennes-94}.
For 3D microemulsions, however, it is known that the effective 
diffusion coefficient varies linearly with $q$ for large wave 
numbers~\cite{nonomura-99}.

Although the analogy between a 3D and 2D microemulsion has been invoked,
it should be pointed out that a 3D microemulsion 
formed from an oil/water/surfactant mixture arises predominantly due to the 
differences in relative affinity between the components.
For the 2D case, it is the physical interactions  from the 
hydrocarbon chain packing requirements at the ${\rm L_o}$/${\rm L_d}$
interface that gives rise to the lineactant properties of the hybrid lipid.

\section{Discussion}
\label{discussion}

In summary, we have calculated the decay rate of concentration
fluctuations in a  multicomponent fluid membrane for two geometries.
First we considered the membrane surrounded by solvent of finite
depth which is further bounded by two walls.
In the second geometry, we allow the solvent depth on one side of the 
membrane to be infinitely large. 
This is equivalent to a supported membrane in experimental situations.
The resulting integrals for the effective diffusion coefficient of the
concentration fluctuations are calculated numerically to determine the 
various dependencies.
We have also explored a possibility of considering the multicomponent
membrane as a 2D microemulsion.

The present work follows Seki {\it et al.} who were able to use 
analytical means to obtain the effective diffusion 
coefficient~\cite{seki-07}.
Although their result should be valid only in the limit of very small 
$h$, the agreement between the numerical result and the analytical 
expression is fairly well as long as $h \leq \nu^{-1}$.
The opposite limit of very large $h$ was studied by Inaura and 
Fujitani through numerical methods~\cite{inaura-08}.
Using the general mobility tensor given by eqn~(\ref{eqn:genoseen}),
we are able to probe all the intermediate situations of finite $h$.
We have verified that our results properly interpolates between
these previous works in the limits of $H\to0$ and $H\to\infty$.

In our case, the measure of the size of the transient structures 
is given by the correlation length $\xi$.
Close to the critical point, $\xi$ diverges and the 
hydrodynamic effects of the outer fluid play a significant role in 
altering the diffusion coefficient.
It has been previously shown through explicit calculations that the 
diffusion coefficient has a logarithmic behavior $D \approx \ln(1/R)$
when the size of the diffusing object $R$ is less than the hydrodynamic 
screening length, i.e., $R \ll \nu^{-1}$~\cite{saffman-75,saffman-76}.
When $R \gg \nu^{-1}$, on the other hand, the fluid flow in the bulk 
leads to the diffusion coefficient to show $1/R$-behavior 
(this condition occurs when there are no walls)~\cite{hughes-81}.
In the presence of walls or a substrate, the screening length is altered 
to $\sqrt{h/\nu}$, where $h$ is the distance of the walls from the 
membrane~\cite{stone-98}.
When $R \gg \sqrt{h/\nu}$, the diffusion coefficient now shows an
algebraic decay $D\approx 1/R^2$~\cite{seki-07}.
This change is attributed to the loss of momentum from the membrane to 
the walls whereas momentum is conserved otherwise~\cite{diamant-09b}.

In the present study, we have kept the model as simple as possible.
The bending stiffness of typical membranes is of the order of 
$10 k_{\rm B}T$ which is sufficiently large enough to neglect the 
out-of-plane displacements of the membrane itself.
The dynamics of the out-of-plane fluctuations have been previously 
studied by Levine and MacKintosh~\cite{levine-02}.
In our treatment, we have also neglected the effects of membrane 
curvature which can be significant when the radius of curvature 
becomes close to the hydrodynamic screening length.
Recent calculations by Henle {\it et al.} considered the diffusion 
of a point object on a spherically closed 
membrane~\cite{henle-08,henle-10}.
The extension of this work to finite sized objects is a particularly 
difficult proposition.

From the experiments on model multicomponent vesicles, the 
static critical exponents for the order parameter and correlation 
length were found to have values close to $\bar\beta=1/8$ and 
$\bar\nu=1$, respectively~\cite{hsmith-08}.
Furthermore, experiments on giant plasma membrane vesicles measured 
the critical exponent $\bar\gamma=7/4$ which characterizes the critical
behavior of the osmotic compressibility~\cite{veatch-08}.
These static exponents coincide with the exact results of the 2D Ising 
model~\cite{rowlinson-widom,goldenfeld}.
The description presented in this paper uses the mean-field approach
and therefore the corresponding static exponents are $\bar\beta=1/2$,
$\bar\nu=1/2$, and $\bar\gamma=1$, respectively. 
However, attributing biological relevance to these critical fluctuations
should carried out with caution.
This is because the plasma membranes do not show phase separation
phenomena without chemical treatment~\cite{kaiser-09}.
Extraction of membranes from real cells also ruptures the association with
the underlying cytoskeleton and other active cellular processes like 
vesicular trafficking.

As mentioned earlier, experiments on real plasma membranes have 
suggested that the cell maintains the membranes at a critical 
composition~\cite{veatch-08}.
This leads to nanometer-sized composition fluctuations at the
physiological temperatures, although these structures are much 
smaller than what can be resolved through optical microscopy.
We therefore speculate that there is some biological relevance
in studying concentration fluctuations.
However, this article has mainly concentrated on the dynamics 
towards the equilibrium state of lipid bilayer membranes. 
In real cells, there are many \text{active} non-equilibrium cellular 
processes that are involved in the proper functioning of the cells.
It has indeed been proposed that nano-domain formation may be 
related to the underlying cytoskeleton~\cite{mayor-04}.
There have been several other models which make use of the active
non-equilibrium phenomena to explain the existence of finite sized 
domains in multicomponent membranes.
A review of the non-equilibrium models can be found in 
the review article~\cite{fan-10c}.

Another experimental system which can be used to verify our model is the 
Langmuir monolayer setup.
The upper ``+'' region in a Langmuir  monolayer system is occupied by air
and hence $\eta_{\rm s}^+$ should be set to zero in 
eqn~(\ref{eqn:genoseen}).
In this case, the mobility tensor is given by eqn~(\ref{eqn:cothoseen}),
but the definition should be replaced with $\nu=\eta_{\rm s}^-/\eta$.
The diffusion coefficient as a function of wave vector can be obtained
via light scattering techniques.

\begin{acknowledgments}

We thank H. Diamant, Y. Fujitani, T. Kato and N. Oppenheimer 
for useful discussions.
This work was supported by KAKENHI (Grant-in-Aid for Scientific
Research) on Priority Area ``Soft Matter Physics'' and Grant
No.\ 21540420 from the Ministry of Education, Culture, Sports, 
Science and Technology of Japan.

\end{acknowledgments}


%

\end{document}